\UseRawInputEncoding
\documentclass[10pt,conference]{IEEEtran}
\usepackage{multirow}
\usepackage{cite}
\usepackage{psfrag}
\usepackage{epstopdf}
\usepackage{multicol}
\usepackage{graphicx}
\usepackage{mathtools}
\usepackage{commath}
\usepackage{amsmath}
\usepackage{graphicx}
\DeclareMathSizes{10}{9}{8}{8}
\usepackage{algorithm}
\usepackage{algpseudocode}
\usepackage{algorithmicx}
\usepackage{epstopdf}
\usepackage{multirow}
\IEEEoverridecommandlockouts
\usepackage{times}

\usepackage{multirow}
\usepackage{setspace}

\usepackage{mathtools}
\usepackage{commath}
\usepackage{amsmath}
\usepackage{amssymb}

\usepackage{graphicx}

\usepackage{amsmath}
\usepackage{amsfonts}
\usepackage{mathtools}
\usepackage{commath}
\usepackage{algorithm}
\usepackage{algpseudocode}
\usepackage{algorithmicx}
\usepackage{cite}

\usepackage{float}
\begin{document}
\title {Single-channel speech enhancement by using psychoacoustical model inspired fusion framework}
\author{\IEEEauthorblockN{Suman Samui} 
\IEEEauthorblockA{Department of Electronics and Telecommunication Engineering,\\ Indian Institute of Engineering  Science and Technology Shibpur, Howrah, 711103, India\\ 
Email: samuisuman@gmail.com}}
\maketitle

\begin{abstract}
When the parameters of  Bayesian Short-time Spectral Amplitude (STSA) estimator for speech enhancement are selected based on the characteristics of the human auditory system, the gain function of the estimator becomes more flexible. Although this type of estimator in acoustic domain is quite effective in reducing the back-ground noise at high frequencies, it produces more speech distortions, which make the high-frequency contents of the speech such as friciatives less perceptible in heavy noise conditions, resulting in intelligibility reduction. On the other hand, the speech enhancement scheme, which exploits the psychoacoustic evidence of frequency selectivity in the modulation domain, is found to be able to increase the intelligibility of noisy speech by a substantial amount, but also suffers from the temporal slurring problem due to its essential design constraint. In order to achieve the joint improvements in both the perceived speech quality and intelligibility, we proposed and investigated a fusion framework by combining the merits of acoustic and modulation domain approaches while avoiding their respective weaknesses. Objective measure evaluation shows that the proposed speech enhancement fusion framework can provide consistent improvements in the perceived speech quality and intelligibility  across different SNR levels in various noise conditions, while compared to the other baseline techniques.  

\end{abstract}

\section{Introduction}
Speech enhancement covers a broad spectrum of applications and objectives, ranging from assistive listening devices to mobile communication. Signal processing solutions to the speech enhancement problem therefore have been approached from various perspectives. The main objective of speech enhancement is to reduce the corrupting noise component (improving quality) of a noisy speech signal while preserving the intelligibility of the original clean speech  as much as possible [1]. However, none of the well-known techniques of speech enhancement was found to be promising in improving the speech intelligibility relative to unprocessed corrupted speech \cite{hu2007}. Moreover, these techniques increase the perceived quality at the expense of reduced intelligibility by introducing distortions to the original speech signal, and residual noise, sometimes in the form
of annoying artefacts known as $musical$ $noise$ in the processed speech signal \cite{Loizou2013}. Although, a large number of research papers on different approaches and methods have managed to address these problems with varying degrees of success, the perfect solution seems to be quite elusive \cite{parchami2016recent} \cite{Loizou2011}. Recently,  deep neural network based data-driven speech enhancement systems have shown enormous potential in improving speech quality and intelligibility, but the generalization of these systems in different dimensions (SNRs, noise and speaker) is still an open issue and requires more research attention \cite{sie2017}.

In this work, we have investigated the effectiveness of pyschoacoustics in both the acoustic and modulation domain  for single-channel speech enhancement (SCSE) task. We derived the gain function of a parametric Bayesian STSA estimator under the Generalized Gamma Distribution (GGD) speech prior assumption. The parameters of this estimator are chosen based on the human auditory characterestics such as cochlea's dynamic compressive non-linearity or loudness perception theory.  This psychacoustic dependence of parameters provide more flexibility in the gain function of STSA estimator. It makes the estimator also more effective in reducing the noise at high frequencies while limiting the speech distortions at lower frequencies. On the other hand, by being motivated by psychoacoustic evidence of frequency selectivity in the modulation domain, we explored a  modulation domain binary masking scheme which retains only speech dominated modulation components in the low modulation frequency (2-16 Hz) based on the modulation domain SNR criterion \cite{dubbelboer2008concept}. This method is able to increase the intelligibility of noisy speech by a substantial amount, but also suffers from the temporal slurring problem. In order to further achieve the joint improvements in both the perceived speech quality and intelligibility, we proposed a fusion framework by combining the merits of acoustic and modulation domain approaches while avoiding their respective weaknesses. The fusion is performed in the short-time spectral domain by combining the magnitude spectra of the above speech enhancement algorithms. Objective evaluation of the speech enhancement fusion shows consistent improvement in the speech quality and intelligibility across different input SNR levels.

The remainder of the paper is organized as follows. In Section 2, we derived the gain function of weighted $\beta$-order estimator and show how the parameters can be selected based on the human auditory characteristics. Section 3 presents modulation domain processing scheme. Next, the proposed fusion framework is presented in Section 4. Experimental results are described in Section 5. Finally, Section 6 concludes the work. 


\section{Signal model and notation}
At the input, single-channel speech enhancement system only observe the noisy speech signal, $y(n)$,  which consists of a clean speech, $s(n)$, and the additive noise signal, $w(n)$, that is statistically independent of $s(n)$. The sampled $y(n)$ is split into overlapping segments and each segment is transformed to the Fourier domain after an analysis  window has been applied.  We can assume that the complex STFT coefficients of the noisy speech $Y(p,k)$ are given by an additive superposition of uncorrelated zero-mean speech coefficients  $S(p,k)$ and noise coefficients $W(p,k)$ as
\begin{equation}
Y(p,k) = S(p,k) + W(p,k)
\end{equation}
where $p$ and $k$ denote the acoustic frame index and the acoustic frequency index respectively. 
The frame index and frequency index shall be discarded for better readability. 
\section{Gain function ($G_A$) of the parametric Bayesian STSA estimator}

\subsection{Derivation of gain function}
The Bayesian STSA estimation problem can be formulated
as the minimization of the expectation of a cost
function $C(A,\hat{A})$ that represents a measure of distance between the true and estimated speech STSAs, denoted respectively
by $A$ and $\hat{A}$. 
The optimal speech STSA estimate in a Bayesian sense can be expressed as 
\begin{equation}
\begin{split}
\hat{{A}}^{(o)} &= \underset{\hat{A}} {\mathrm{arg min}} \hspace{2 mm} E[C(A,\hat{A})]\\
&=\int \bigg{[} \int C(A,\hat{A})p(A|Y)dA\bigg{]}p(Y)dY
\end{split}
\end{equation}
where $p(A,Y)$ and $p(A|Y)$ being the joint and conditional
PDFs of the speech STSA and complex STFT coefficients $Y$ respectively. 

In this work, we have considered a parametric Bayesian estimator whose cost function is given by
\begin{equation}
C(A,\hat{A}) = \bigg{(}\dfrac{A^{\beta}-\hat{A}^{\beta}}{A^{\alpha}}\bigg{)}^{2}
\end{equation}
where $\alpha$ and $\beta$ are two adaptive parameters.  Substituting (4) into (3) and minimizing the expectation, the following is obtained
\begin{equation}
\hat{{A}}^{(o)}=\Bigg{(}\dfrac{E\{A^{\beta-2\alpha}|Y\}}{E\{A^{-2\alpha}|Y\}}\Bigg{)}^{1/\beta}
\end{equation}
The conditional moments of the form $E\{A^{M}|Y\}$ appearing
in (5) can be obtained as
\begin{equation}
E\{A^{M}|Y\}=\dfrac{\int_{0}^{\infty} \int_{0}^{2\pi} A^{M}p(Y|A,\theta_s)p(A,\theta_s)d{\theta_s}dA}{\int_{0}^{\infty} \int_{0}^{2\pi}p(Y|A,\theta_s)p(A,\theta_s)d{\theta_s}dA}
\end{equation}
with $p(Y|A,\theta_s)$ and $p(A,\theta_s)$ being respectively the conditional PDF of the noisy observation given the clean speech
and the joint PDF for the speech amplitude and phase.

 In this work, we explore the use of generalized Gamma distributed (GGD) speech STSA priors, which are experimentally shown \cite{shin2005statistical} to more accurately approximate empirical histograms of speech ( particularly when the frame-size is less than 100 ms \cite{martin2005speech}),
\begin{equation}
p(A) = \frac{\kappa \lambda^\mu}{\Gamma(\mu)}A^{\kappa\mu-1}\exp(-{\lambda}A^{\mu})
\end{equation}
where $\kappa$ and $\mu$  are known as the shape parameters and $\lambda$ as the scaling parameter \cite{borgstrom2011unified}. $\Gamma(.)$ denotes Gamma function. In order to get a close-form solution, we have set $\kappa$ = 2 which makes (7) to a generalized form of $\chi$-distribution. Based on the second moment of the derived $\chi$-distribution, it can be deduced that the two parameters $\lambda$ and $\mu$ must satisfy the relation $\mu$/$\lambda$ = ${\sigma_s^2}$ \cite{borgstrom2011unified}. 
Now assuming the uniform PDF for the speech spectral phase and complex zero-mean Gaussian PDF for the noise spectral
coefficients, we can derive the gain function of the STSA estimator by using (8), (6) and (5),
\begin{equation}
G = \sqrt{\frac{\gamma}{\frac{\mu}{\zeta}+1}}\Bigg{(}\dfrac{\Gamma(\frac{\beta}{2}+\mu-\alpha)M(\frac{2-\beta}{2}+\alpha-\mu,1;-\nu)}{\Gamma(\mu-\alpha)M(1+\alpha-\mu,1;-\nu)} \Bigg{)}^{1/\beta}
\end{equation}
where $\gamma$ = $\dfrac{R^2}{E\{V^2\}}$= $\dfrac{R^2}{{\sigma_w^2}}$, $\zeta$ = $\dfrac{E\{A^2\}}{E\{V^2\}}$= $\dfrac{{\sigma_s^2}}{{\sigma_w^2}}$ and $\nu$ = $\dfrac{\zeta}{\mu+\zeta}\gamma$. The parameters $\zeta$ and $\gamma$ are called the $a$ $priori$ and $a$ $posteriori$ SNRs, respectively. $M(.)$ denotes the confluent hypergeometric functions. As observed in Figure, increasing the shape parameter $\mu$ leads to a monotonic increase of the gain function for all considered values of SNR, we proposed to choose $\mu$ as a linear function of SNR at each time frame as
\begin{equation}
\mu(p) = \mu_{min} + (\mu_{max}-\mu_{min})\zeta_{norm}(p) 
\end{equation}
where, based on the comprehensive experimentations, $\mu_{min}$ and $\mu_{max}$ are chosen as 1 and 3, respectively and $\zeta_{norm}$ denotes the normalized $a$ $priori$ SNR of frame $p$.
\subsection{Selection of parameters: $\alpha$ and $\beta$}
The value of $\alpha$ can be chosen by taking advantages of the masking properties of human ear. It is always desirable for a STSA estimator to favor a more accurate estimation of smaller
STSA since they are less likely to mask noise remaining in the
clean speech estimate. Since most of the speech energy is located at lower frequencies, higher frequencies should contain mainly small STSA \cite{formby1982long}. Therefore, it would be relevant to further increase the weights of the smaller STSA in the cost function for higher frequencies. This can be done by increasing $\alpha$  for higher frequencies as follows
\begin{equation}
{\alpha}_{k} = \begin{cases} {\alpha}_{low}, & {f}_{k} \leq 2kHz \\ \dfrac{({f}_{k}-2000)({\alpha}_{high}-{\alpha}_{low})}{\frac{{f}_{s}}{2}-2000}+{\alpha}_{low}, & otherwise \end{cases}
\end{equation}
where parameters, ${\alpha}_{high}$ and ${\alpha}_{low}$ are set empirically. On the other hand, by considering the non-linearity in dynamic range compression in the perception of loudness in the human cochlea, we can select the $\beta$ value as follows,
\begin{equation}
{\beta}_{k} = \Bigg{[}\dfrac{\log_{10}(\frac{f_k}{Q}+l)}{\log_{10}(\frac{f_s}{2Q}+l)}\Bigg{]}({\beta}_{high}-{\beta}_{low}) +  {\beta}_{low}
\end{equation}
where $Q$ = 16.54 is an empirical constant relevant to the $tonotopic$ $mapping$ of basilar membrane. 


 \begin{figure*}[t]

 \centering

  \begin{tabular}{ccc}


    \includegraphics[scale = 0.3]{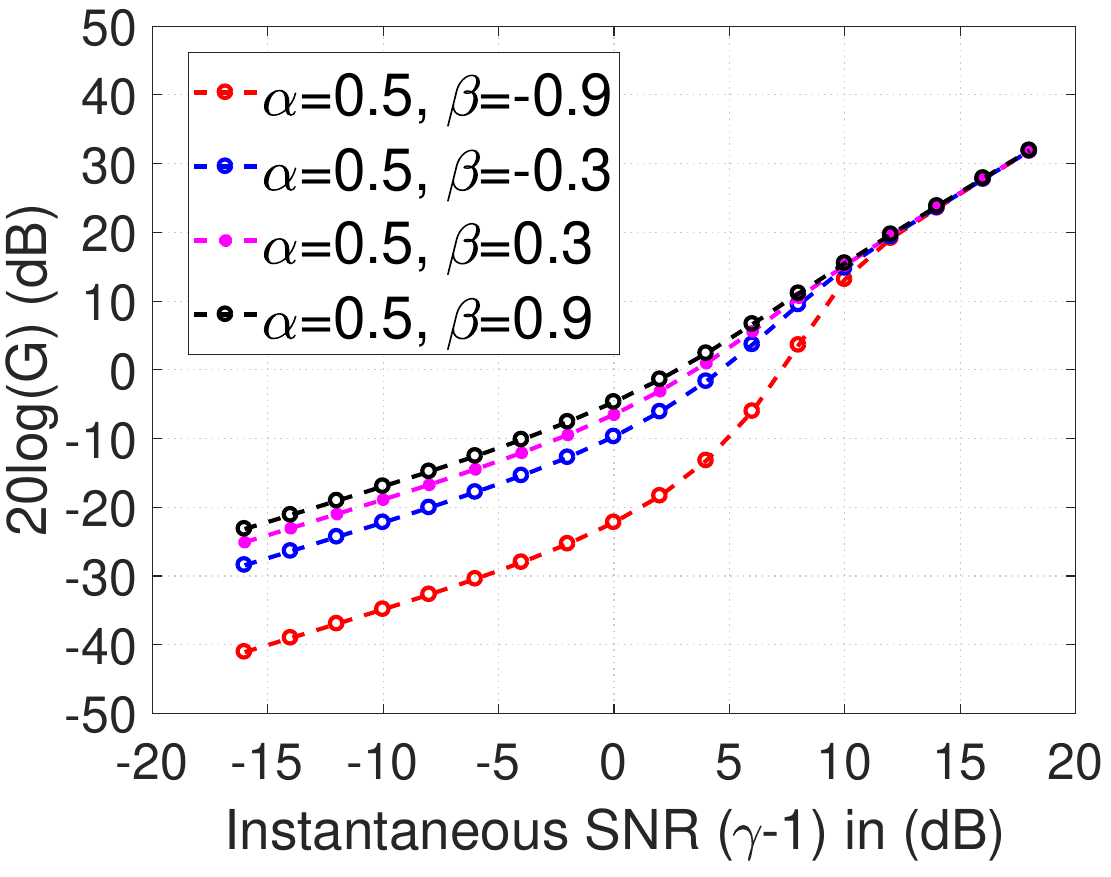} &
    
    \includegraphics[scale = 0.3]{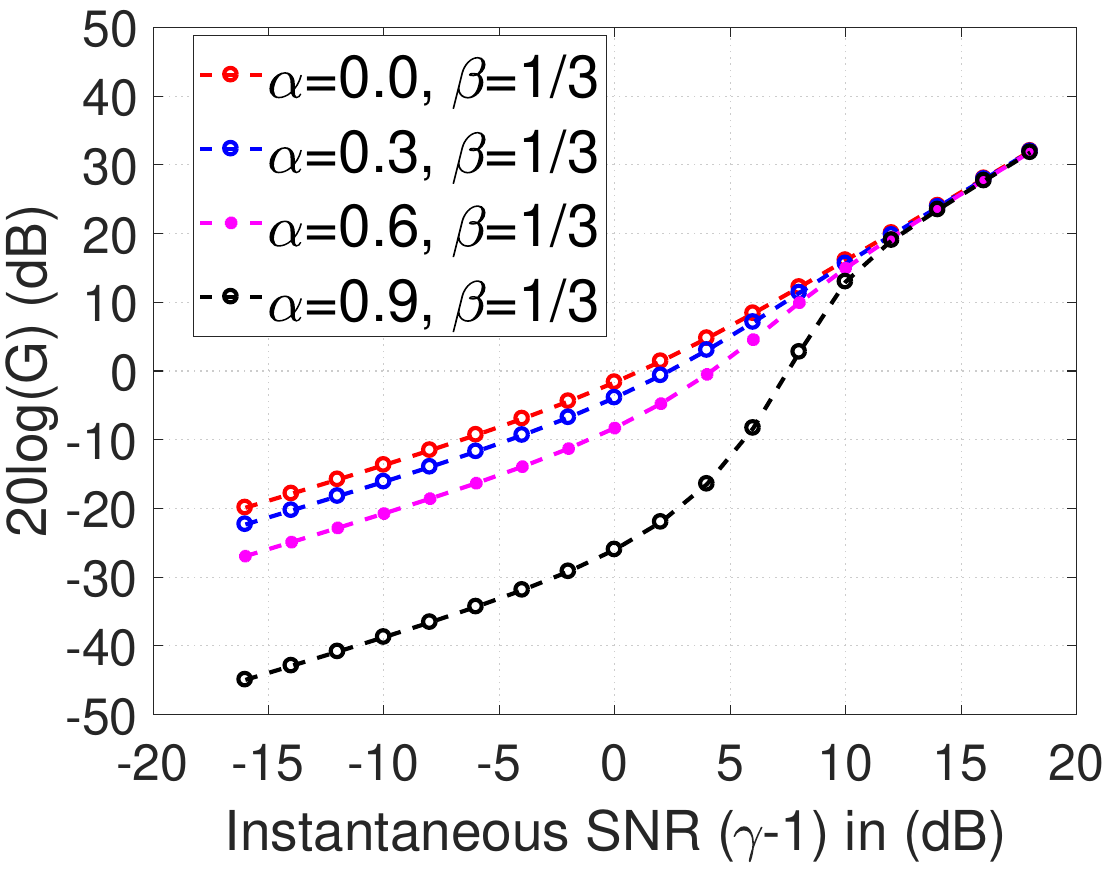} &
     
    \includegraphics[scale = 0.3]{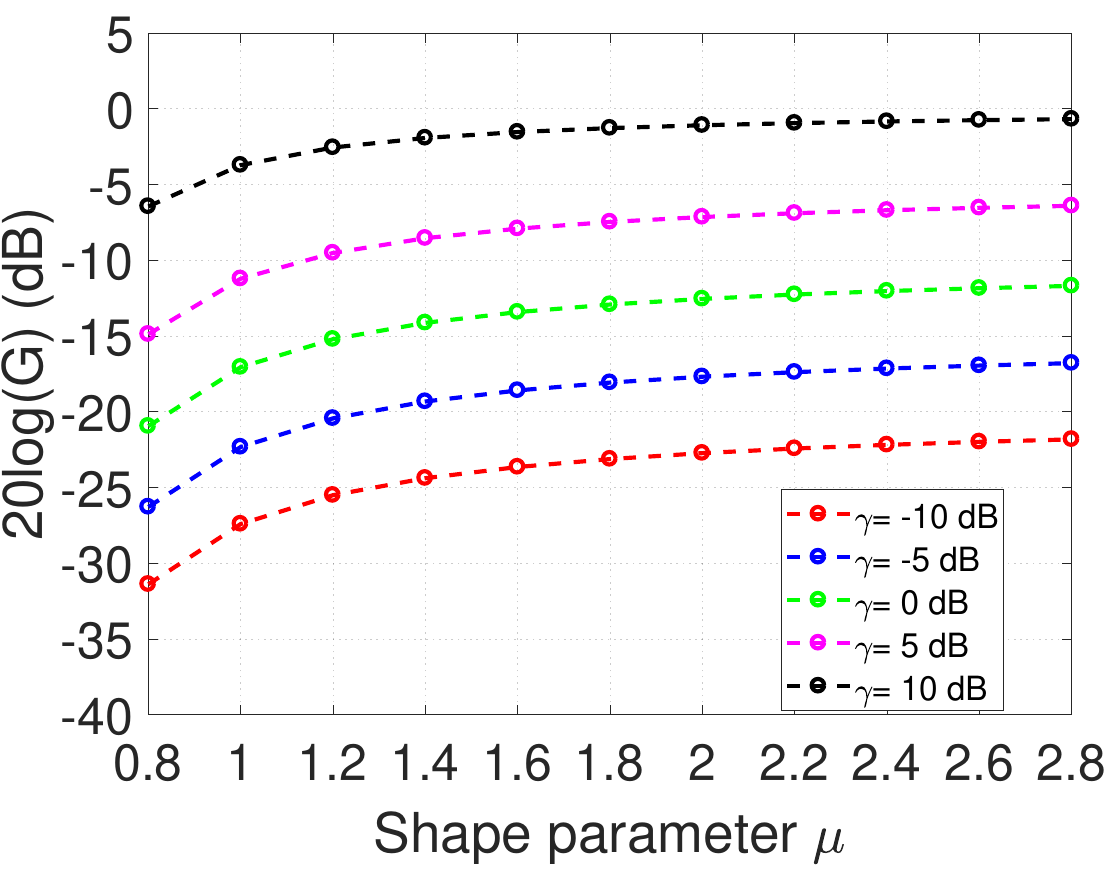} \\

    (a) \ & (b) & (c) \\

  \end{tabular}
  \label{fig4}\caption{(a) Gain of the parametric STSA estimator $(20\log G)$ versus instantaneous SNR $(\gamma-1)$ for several values of $\beta$. (b) $20\log G$ versus $(\gamma-1)$ for several values of $\alpha$ (c) $20\log G$ versus $(\gamma-1)$ for several values of $\mu$  for $\zeta$ = 0 dB }
\end{figure*}

%
%
%
%
%
%
%
%
%
%
%
%
%
%
%
%
%
\begin{figure}[h]
  \centering
  \includegraphics[scale = 0.35]{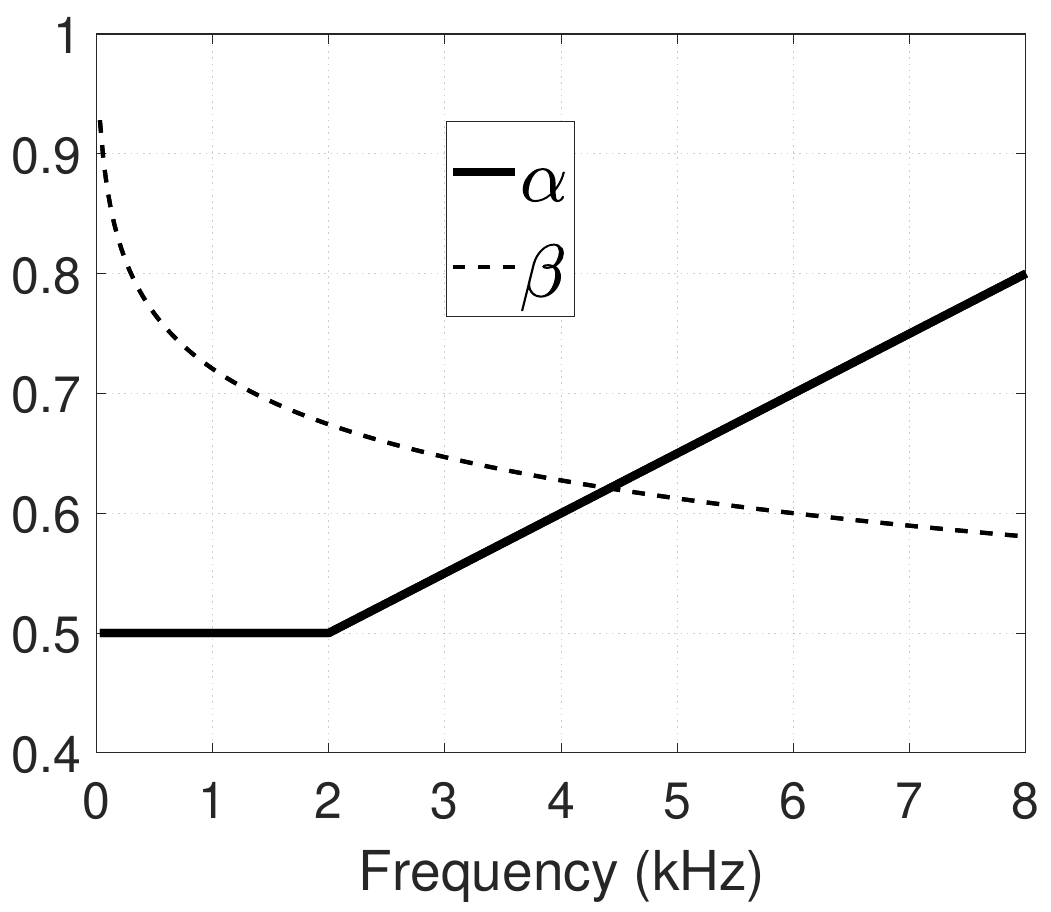}
  \caption{$\alpha$ and $\beta$ variations with frequency}
  \label{fig1}
\end{figure}

\section{Modulation domain processing}
Modulation spectrum can be obtained by applying an another STFT on each of the acoustic frequency track $k$ of the acoustic magnitude spectrum.  In this study, we have retained the speech-dominant low-frequency (2-16 Hz) modulation spectrum components (also referred to as $modulation$ $channels$) while the remaining modulation components are discarded based on  signal-to-noise ratio measured in the modulation domain as a selection criterion \cite{wojcicki2012channel}\cite{dubbelboer2008concept}. This is being motivated by the fact that the intelligible components of the speech signal are mostly confined to the modulation frequency band of $2$Hz to $16$Hz \cite{ewert2000characterizing}\cite{drullman1994effect}. The corresponding binary gain function is embodied in (12) as follows
\begin{equation}
\begin{split}
\mathcal{\tilde{G}}(q,k,m)&= \begin{cases} 1, \: if \: \: {\xi}(q,k,m)\geq \eta_{th}\: and \: m\leq M_c  \\ 0, \: otherwise \end{cases}
\end{split}
\end{equation}
where $q$ denotes the modulation frame index, $k$ is the index of the acoustic frequency, $m$ refers to the index of the modulation
frequency. The parameter $M_{c}$ denotes the modulation cut-off frequency and $\eta_{th}$ is referred to as modulation domain SNR threshold. The modulation domain SNR ${\xi}(q,k,m)$  can be computed as
\begin{equation}
{\xi}(q,k,m)=\dfrac{|\hat{\mathcal{S}}(q,k,m)|^2}{|\hat{\mathcal{V}}(q,k,m)|^2}
\end{equation}
where $|\hat{\mathcal{S}}(q,k,m)|$ is an estimate of clean modulation spectrum, computed using the spectral subtraction method applied in modulation domain and $|\hat{V}(q,k,m))|$  is an estimate of modulation spectrum of noise, computed from the speech-absent portion of the noise masked stimulus. Empirically, we found that the value of  $\eta_{th}$ equal to -$10$ dB gives the best improvement in intelligibility while the $M_c$ is set to 4 Hz. Compared to acoustic domain transformation (first STFT), the large window size is required for the second STFT in order to achieve sufficiently good frequency resolution near 4 Hz in the modulation frequency. 

\begin{table}[]
\centering
\caption{Parameters for STFT analysis in dual-AMS framework}
\label{my-label}
\begin{tabular}{ccc|ccc} \hline
\multicolumn{3}{c|}{{$\mathbf{STFT_a}$}}       & \multicolumn{3}{c}{{$\mathbf{STFT_m}$}}        \\ \hline 
\textbf{Wa} & \textbf{FSa} & \textbf{FFTa} & \textbf{Wm} & \textbf{FSm} & \textbf{FFTm} \\ \hline
32 ms       & 16 ms        & 512           & 256 ms       & 32 ms         & 64    \\ \hline      
\end{tabular}
\end{table}
\begin{table*}[t]
	\centering
	\scriptsize
	\renewcommand{\arraystretch}{1.4}
	\caption{PESQ and ESTOI scores averaged over test data [represented as PESQ(ESTOI)]}
	\label{my-label}
	\begin{tabular}{||c|c|c|c|c|c|c|c|c||}
		\hline \hline
		\textbf{Noise Type} & \textbf{SNR} &
		\textbf{\begin{tabular}[c]{@{}c@{}}Noisy\\ (Unprocessed)\end{tabular}} &
		\textbf{\begin{tabular}[c]{@{}c@{}}Proposed\\ Fusion approach\end{tabular}} &
		\textbf{MMSE-LSA} & \textbf{WE}$(p=-1)$ & \textbf{AMB-STSA} & \textbf{CTSP} & \textbf{Mod-SSub} \\ \hline \hline
		
		\multirow{3}{*}{Babble}  & 0 dB  & 1.65(0.42) & \textbf{2.35(0.63)} & 2.05(0.57) & 1.92(0.53) & 1.98(0.55) & 1.88(0.52) & 2.12(0.58) \\ \cline{2-9}
		& 5 dB  & 2.25(0.55) & \textbf{2.95(0.74)} & 2.62(0.69) & 2.48(0.66) & 2.55(0.67) & 2.44(0.65) & 2.70(0.71) \\ \cline{2-9}
		& 10 dB & 2.95(0.68) & \textbf{3.55(0.83)} & 3.28(0.79) & 3.12(0.76) & 3.20(0.77) & 3.08(0.75) & 3.35(0.80) \\ \hline \hline
		
		\multirow{3}{*}{Pink}    & 0 dB  & 1.85(0.46) & \textbf{2.55(0.66)} & 2.22(0.60) & 2.10(0.58) & 2.16(0.59) & 2.05(0.57) & 2.30(0.62) \\ \cline{2-9}
		& 5 dB  & 2.45(0.58) & \textbf{3.10(0.77)} & 2.82(0.72) & 2.68(0.70) & 2.74(0.71) & 2.62(0.69) & 2.90(0.74) \\ \cline{2-9}
		& 10 dB & 3.15(0.70) & \textbf{3.70(0.85)} & 3.40(0.81) & 3.26(0.79) & 3.32(0.80) & 3.22(0.78) & 3.48(0.82) \\ \hline \hline
		
		\multirow{3}{*}{White}   & 0 dB  & 1.95(0.50) & \textbf{2.70(0.69)} & 2.35(0.63) & 2.20(0.60) & 2.28(0.61) & 2.15(0.59) & 2.42(0.64) \\ \cline{2-9}
		& 5 dB  & 2.60(0.62) & \textbf{3.25(0.80)} & 2.95(0.75) & 2.78(0.72) & 2.86(0.73) & 2.72(0.71) & 3.02(0.76) \\ \cline{2-9}
		& 10 dB & 3.30(0.74) & \textbf{3.85(0.88)} & 3.55(0.84) & 3.38(0.81) & 3.46(0.82) & 3.34(0.80) & 3.60(0.85) \\ \hline \hline
		
	\end{tabular}
\end{table*}

\section{Proposed fusion framework}
The main drawback of the parametric Bayesian STSA estimator (described in Section 2) is that there is always a decrease in the gain $G$ at high frequencies compared to lower frequencies (please refer to Figure). This decrease in $G$ generates more noise reduction at high frequencies but has the simultaneous effect of producing more speech distortions. In the presence of produced distortion, the high frequency contents of speech, such as fricatives, will be less perceptible in heavy noise (i.e. low SNRs), which leads to low intelligibility. On the other hand, the modulation domain channel selection techniques (described in Section 3), can able to increase intelligibility of noisy speech by a substantial amount compared to the acoustic domain Bayesian STSA estimator, but the main disadvantage of this technique is the time smearing (slurring) problem which mainly occurs due to the use of long window length in second STFT analysis of dual-AMS framework. The long modulation analysis window is necessary to obtain a good resolution at the low modulation frequencies. In order to exploit the strengths of the two methods, while trying to avoid their weaknesses, by combining (fusing) them in the acoustic STFT domain, as the following
\begin{equation}
\hat{S}(k,p)=\Phi(\psi)\hat{S}_A(k,p)+(1-\Phi(\psi))\hat{S}_M(k,p)
\end{equation}
where $\Phi$(.) is called the fusion-weighting function. It's value depends on the instantaneous SNR $\psi$ =  $\dfrac{R^2-{\hat{\sigma}_w^2}}{{\hat{\sigma}_w^2}}$ = ($\gamma$-1) in the acoustic domain and can be empirically  set as follows: 
\begin{equation}
\Phi(\psi)= \begin{cases} 0.2, & \psi \leq 2 dB \\ \dfrac{\psi-2}{14}, & 2 dB< \psi < 16 dB \\ 0.8, & \psi \geq 16 dB \end{cases}
\end{equation}
The above weighting favours the modulation domain filtering method at low segment SNRs (i.e., during
speech pauses and low energy speech regions), while stronger
emphasis is given to the acoustic domain Bayesian estimator at high segment SNRs (i.e., during high energy speech regions).

\section{Experimental results}

\subsection{Speech and noise corpus}
To evaluate the performance of the proposed approach, we have synthetically generated noisy speech stimuli by adding noise to thirty clean utterances (Harvard sentences) taken from IEEE corpus. The utterances comprise of 3 male and 3 female speaker ( 5 sentences by each). Three types of noise instances: white, babble and pink noise, are taken from NOISEX-92 database. All the utterances and noise samples are down-sampled to 16 kHz. Each utterance then is mixed with the aforementioned noise instances at three SNR levels (0,5 and 10 dB). To obtain desired SNR, the noise level is adjusted based on `active speech level' according to ITU-T P.56 \cite{ITU-TASL}.

The proposed fusion based speech enhancement approach uses a dual-AMS framework which uses two step STFT analysis. The various parameters (the window size, frame-shift and FFT size) for implementing these STFTs are tabulated in Table 1. The $decision$-$directed$ approach has been employed to compute the $a$ $priori$ SNR in the acoustic domain. Noise PSD has been computed by employing unbiased MMSE based estimator.

\begin{figure*}[h]
\centering
\includegraphics[scale=0.75]{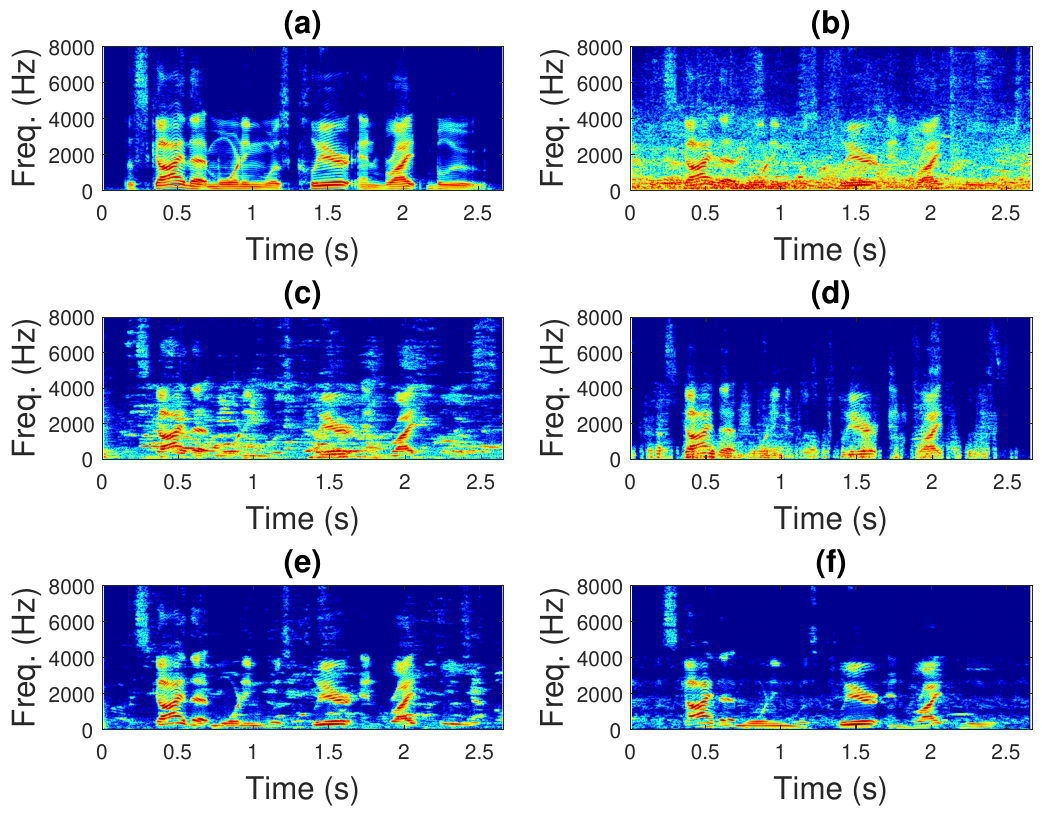}
\caption{Magnitude spectrograms: (a) clean utterance (PESQ: 4.5) (b) speech degraded by babble boise at 0 dB SNR and noisy speech enhanced using (c) MMSE-LSA (d) PMB-STSA (e) MCS (f) the proposed fusion approach.}
\end{figure*}
\label{fig1}
\subsection{Objective measure evaluation}
In the current work, two objective measures PESQ \cite{recommendation2001perceptual} and ESTOI \cite{jensen2016algorithm} are considered. Both of them are considered to be highly correlated with subjective listing test (such as MOS). PESQ provides a score in the range of [1,4.5] and ESTOI in [0,1].    
We compare the performance of the proposed fusion SCSE approach with (i) the auditory model based motivated Bayesian STSA estimator (described in Section 2) denoted by AMB-STSA, (ii) MMSE-LSA estimator \cite{ephraim1985speech}, (iii) combined temporal and spectral domain processing (CTSP) method \cite{krishnamoorthy2011enhancement}, (iv) perceptually motivated Weighted Euclidean (WE) STSA estimator \cite{loizou2005} and (v) Modulation domain spectral subtraction technique (Mod-SSub) \cite{paliwal2010single}. The average comparative result in terms of PESQ and STOI scores of enhanced speech signals for different methods along with noisy speech in various noise conditions are shown in Table 2. 


\section{Conclusions}
In this paper, we have proposed a fusion framework for solving the speech enhancement task. The proposed method combines the advantages of an acoustic domain psychoacoustically motivated parametric Bayesian estimator and low frequency selective modulation domain binary masking based on modulation domain SNR criterion.  Objective measure evaluation confirmed that the proposed method can outperform the other baseline techniques in terms of perceived speech quality and intelligibility metrics.   
    
\bibliographystyle{IEEEtran}
\bibliography{ref}
\end{document}